\documentclass[usenatbib,useAMS,a4paper,fleqn]{mnras}
\usepackage{amssymb}
\usepackage{newtxtext,newtxmath}
\usepackage[T1]{fontenc}
\usepackage{ae,aecompl}
\usepackage{times}
\usepackage{microtype}
\usepackage{verbatim}
\usepackage{graphicx}
\usepackage{tablefootnote}
\date{Accepted XXX. Received YYY; in original form ZZZ}
\pubyear{2024}

\usepackage[multidot]{grffile}

\begin{document}

\label{firstpage}
\pagerange{\pageref{firstpage}--\pageref{lastpage}}

\title[Circumplanetary discs from satellites]
      {The formation of transiting circumplanetary debris discs from the disruption of satellite systems during planet--planet scattering}

\author[Mustill et al.]
       {Alexander J. Mustill$^{1,2}$\thanks{E-mail: alex@astro.lu.se},
         Melvyn B. Davies$^3$,
         Matthew A. Kenworthy$^{4}$\\
         $^1$Lund Observatory, Division of Astrophysics, Department of Physics, Lund University, Box 118, SE-221 00 Lund, Sweden\\
         $^2$Lund Observatory, Department of Astronomy \& Theoretical Physics,
         Lund University, Box 43, SE-221 00 Lund, Sweden\\
         $^3$Centre for Mathematical Sciences, Lund University, Box 118, SE-221 00 Lund, Sweden\\
         $^4$Leiden Observatory, University of Leiden, PO Box 9513, 2300 RA Leiden, The Netherlands
       }

\maketitle

\begin{abstract}
Several stars show deep transits consistent with discs of roughly $1\mathrm{\,R}_\odot$ seen at moderate inclinations, likely surrounding planets on eccentric orbits. We show that this configuration arises naturally as a result of planet--planet scattering when the planets possess satellite systems. Planet--planet scattering explains the orbital eccentricities of the discs' host bodies, while the close encounters during scattering lead to the exchange of satellites between planets and/or their destabilisation. This leads to collisions between satellites and their tidal disruption close to the planet. Both of these events lead to large quantities of debris being produced, which in time will settle into a disc such as those observed. The mass of debris required is comparable to a Ceres-sized satellite. Through $N$-body simulations of planets with clones of the Galilean satellite system undergoing scattering, we show that 90\% of planets undergoing scattering will possess debris from satellite destruction. Extrapolating to smaller numbers of satellites suggests that tens of percent of such planets should still possess circumplanetary debris discs. The debris trails arising from these events are often tilted at tens of degrees to the planetary orbit, consistent with the  inclinations of the observed discs. Disruption of satellite systems during scattering thus simultaneously explains the existence of debris, the tilt of the discs, and the eccentricity of the planets they orbit.
\end{abstract}


\section{Introduction}

\label{sec:intro}

%\textcolor{white}{Lorem ipsum this is whitespace to fix a really annoying hyperref error}

Opulent systems of rings and satellites\footnote{Or moons, or, if in other planetary systems, exomoons.} are a feature of all four known giant  planets of our Solar System, and there is now growing evidence of such systems orbiting extra-Solar planets. In 2007, the star 1SWASP J140747.93-394542.6 underwent a long-duration dimming event whose complex light curve likely results from the eclipse of the star by a substellar object hosting an extensive ring system, possibly sculpted by satellites \citep{Mamajek+12,vanWerkoven+14,KenworthyMamajek15,Kenworthy+15,RiederKenworthy16}. The unexpectedly bright and blue object Fomalhaut~b has been interpreted as a planet surrounded by large quantities of dust, either configured as a disc \citep{Kalas+08} or arising from collisions among an irregular satellite swarm \citep{KennedyWyatt11}. An optically-thin disc has been proposed to explain the colour-dependent transit depth of K2-33b \citep{Ohno+22}, which is significantly deeper in the visible than in the IR. Rings may cause anomalies in transit light curves \citep{BarnesFortney04} or affect the inference of planetary density \citep{Zuluaga+15}, and the extremely low density of HIP~41378f \citep{Santerne+19} may be explained if the planet is surrounded by a large ring system \citep{Akinsanmi+20,PiroVissapragada20,Harada+23,Saillenfest+23}. ALMA imaging of the PDS~70 system shows that circumplanetary dust discs can exist at very young ages (few Myr) while the planet itself is still forming in the circumstellar disc \citep{Isella+19}; discs or rings around older objects may be the remnants of these birth discs or alternatively formed later from the disruption of large bodies such as satellites.

In this paper, we explain the origin of a class of large circumplanetary discs discovered in transit around their host stars, a class represented by the objects EPIC~204376071 \citep[henceforth EPIC~2043][]{Rappaport+19}, V928~Tau \citep{vanDam+20}, and EPIC~220208795 \citep[henceforth EPIC~2202][]{vanderKamp+22}. These discs have a large size ($\sim1\mathrm{\,R}_\odot$) and appear to surround planets on eccentric orbits around their host stars, as described in Section~\ref{sec:prelim}. We show that these discs, as well as the orbital eccentricity of their host planets, are easily produced as a result of planet--planet scattering between giant planets hosting satellite systems. 

Planet--planet scattering is thought to be a common occurrence in systems formed with multiple gas giants, and the eccentricity distribution of gas giants is consistent with scattering having occurred in the majority of systems \citep{JuricTremaine08,Raymond+11}. Such scattering, involving often multiple close approaches between different planets, will have dynamical effects on the satellites hosted by these planets. The current direct evidence for such satellites is less strong than that for circumplanetary rings or discs, with to date two candidates detected in transit: Kepler-1625b~I \citep{Teachey+18,TeacheyKipping18} and Kepler-1708b~I \citep{Kipping+22}, the former of which has been disputed \citep{Rodenbeck+18,Heller+19,Kreidberg+19}. On a population level, by stacking transits of 284 KOIs, \cite{Teachey+18} placed a limit of $<38\%$ of planets between 0.1 and 1\,au hosting satellite systems similar to the Galilean satellite system of Jupiter, at 95\% confidence. Constraints on the frequency of planets hosting satellites in the habitable zone, however, were not meaningful ($<97\%$), as fewer transiting planets are known and fewer transits can be stacked. It is possible, therefore, that many giant planets orbiting beyond $\sim1$\,au host satellite systems similar to the gas giants of our Solar System.

%\emph{Current observational status on exomoons, rings and circumplanetary discs}
%\begin{itemize}
%    \item \emph{J1407b: complex occultation light curve interpreted as large ring system/residual circumplanetary disc being sculpted by an exomoon system. \cite{Mamajek+12,vanWerkoven+14,KenworthyMamajek15,Kenworthy+15,RiederKenworthy16}}
%    \item \emph{Other occulters, likely large ring systems: EPIC~220208795 \citep{vanderKamp+22}, EPIC~204376071 \citep{Rappaport+19}, V928~Tau \citep{vanDam+20}. Latter is a binary (projected separation 32\,au) with a tertiary companion at 2\,250\,au.}
%    \item \emph{K2-33b, a young (10Myr) planet possibly with an optically-thin circumplanetary dust disc \citep[explains why transit depth smaller in IR than Vis][]{Ohno+22}.}
%    \item \emph{Exomoon candidates: Kepler-1625b~I \citep{Teachey+18,TeacheyKipping18}. Disputed \citep{Rodenbeck+18,Heller+19,Kreidberg+19}. Kepler-1708b~I: $2.6\mathrm{\,R}_\oplus$, 1.6\,au Jupiter-sized host \citep{Kipping+22}. \cite{Teachey+18}: Most giant planets at $0.1-1$\,au do not have moons but constraints on giants in the HZ are pretty weak.}
%    \item \emph{ALMA CPD results...}
%    \item \emph{Fomalhaut b could be disc/ring system \citep{Kalas+08} or collisions among a satellite swarm \citep{KennedyWyatt11}.}
%    \item \emph{\cite{Kleisioti+23}: prospects for detectability of tidally-heated moons in thermal IR}
%    \item \emph{HIP 41378f: very low-density, long-period planet. Possible     ring system \citep{Harada+23}}
%\end{itemize}

Studies of satellite dynamics during planet--planet scattering in the Solar System have shown that dynamical excitation, collision, ejection or exchange of satellites between planets can result, depending on the proximity of the encounter between planets \citep{Deienno+14,Nesvorny+14,Li+20,LiChristou20}. Such flyby interactions are better studied for the case of stars exchanging or perturbing planets; in particular, instability in the system is not always instantaneous during or immediately after the flyby, but can occur much later, as exchanged planets are often implanted on eccentric, inclined orbits, while the eccentricities of existing planets can also be strongly excited, both of which can seed later instability \citep{Malmberg+11,Hao+13,Li+19,Li+20exo}. Scattering encounters between planets hosting satellites resemble this process in miniature, with the complication that often planets undergo several close encounters as the scattering dynamics resolves, complicating the long-term evolution still further. However, the impact of planet--planet scattering on extra-Solar satellite systems remains comparatively under-studied, though works have shown that tens of percent of satellites can be lost due to ejection from the planetary Hill sphere, collision with another satellite or planet, or tidal disruption \citep{Gong+13,Payne+16,Hong+18,RabagoSteffen19,Trierweiler+22}, depending on the details of the scattering and the satellite orbits. With few exceptions \citep[\emph{e.g.,}][]{Gong+13,RabagoSteffen19} these studies have used test particles to represent the satellites. While computationally efficient, this misses any orbital evolution owing to satellite--satellite interactions, as well as collisions between the satellites.

\begin{figure}
\includegraphics[width=0.5\textwidth]{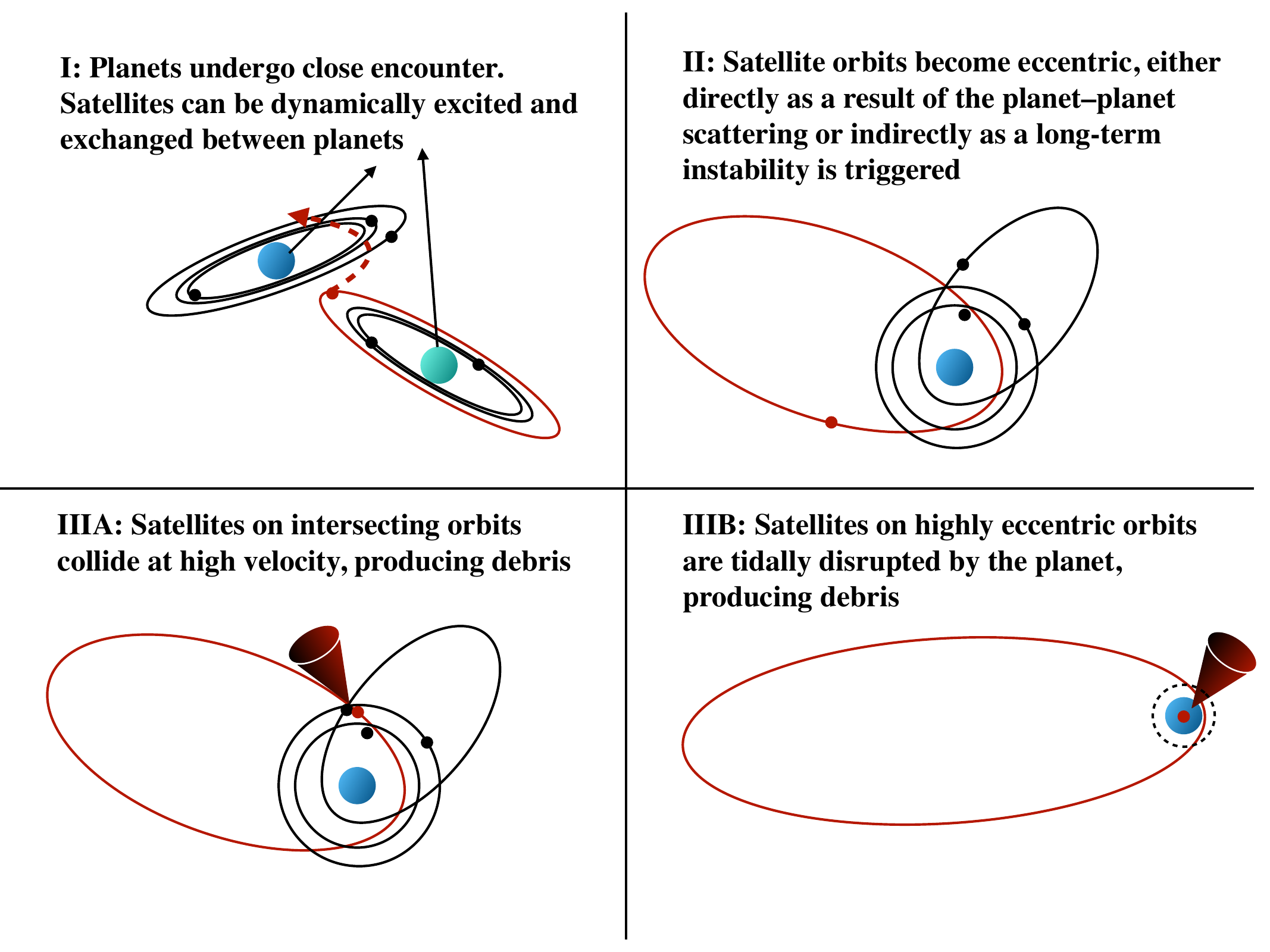}
\caption{Cartoon illustration of the destabilisation and exchange of satellites between planets, and resulting production of debris, through collisions or tidal disruption, that can later settle into a large disc.}
\label{fig:cartoon}
\end{figure}

In this Paper we perform $N$-body simulations of planets undergoing scattering, where the planets possess copies of the Galilean satellite system; we take this, in the absence of any observational data, as a template for extra-Solar satellite systems orbiting gas giants. The satellites are full, massive particles in the simulations, allowing us to adequately capture satellite--satellite dynamics and collisions, which can lead to disc formation along with the tidal disruption of a satellite by the planet. This process is illustrated in Fig~\ref{fig:cartoon}. We discuss the observational background and preliminary calculations in \S\ref{sec:prelim}, describe the $N$-body setup in \S\ref{sec:method}, present results in \S\ref{sec:results}, and conclude in \S\ref{sec:discussion}.

\section{Preliminaries}

\label{sec:prelim}

\subsection{Observations of large occulting discs}

In this paper we focus on a class of system comprising the three stars EPIC~2043, V928~Tau and EPIC~2202. Each of these stars showed a deep (tens of percent), asymmetric dip in its K2 light curve \citep{Rappaport+19,vanDam+20,vanderKamp+22}. The shape and depth of these dips is consistent with the transit of the star by an opaque ellipse, angled with respect to the transit chord. In terms of 3D geometry, this corresponds to the projection of an opaque circle or disc, whose normal is inclined both to the line of sight and to the orbital normal, and  seen in projection as an ellipse. We focus in this paper on EPIC~2202, although all three systems show qualitatively similar properties, and now briefly summarise the model conclusions of \cite{vanderKamp+22} for this system. The star EPIC~2202 has a radius $R_\star=0.83\mathrm{\,R_\odot}$ and a mass of $M_\star=0.85\mathrm{\,M_\odot}$. The inferred properties of the disc are a radius $R_\mathrm{disc}=1.1\mathrm{\,R_\odot}$ and an inclination to the line of sight of $77^\circ$, while the long axis of the projected ellipse is tilted by $37^\circ$ with respect to the transit chord. This does not change significantly between a fully opaque and a ``soft-edged'' disc. The duration of the dip implies a velocity in the plane of the sky of the transiting object across the stellar disc of $77\mathrm{\,km\,s}^{-1}$; this, coupled with the lack of a second transit in the K2 light curve, implies an orbital period $P>60$\,d and an orbital eccentricity $e>0.36$. Supplementing the K2 data with TESS and ground-based photometry yielded a best-fit orbital period of $P=290$\,d and an eccentricity of $e=0.72$. Finally, for the disc to remain bound within the Hill sphere of the object at its centre, that object must have a mass $M_\mathrm{p}>1.5\mathrm{\,M_J}$, \emph{i.e.,} it must be a gas giant planet or brown dwarf.

\subsection{Preliminary calculations}

The transiting objects described above share two 
common features:
\begin{itemize}
    \item A circumstellar orbit which is at least moderately eccentric ($e\gtrsim0.3$).
    \item A large, optically-thick disc, moderately 
    inclined to the line of sight and to the transit chord.
\end{itemize}
A parsimonious explanation for these objects' origin would explain both of these features together. We argue that planet--planet scattering of planets with moons does just this (Fig~\ref{fig:cartoon}).

\subsubsection{Planetary eccentricities from planet--planet scattering}

The broad eccentricity distribution of giant planets suggests that many systems containing multiple giant planets undergo planet--planet scattering early in their history: sufficiently closely-spaced planets initially on near-circular, near-coplanar orbits will see their orbital elements diffuse under the chaotic forcing of mean-motion resonances, leading to the intersection of orbits and close encounters between planets \citep{Wisdom80,Quillen11,Petit+20}. The outcome of a close encounter is determined by the distance of the closest approach: the closer the approach, the stronger the scattering, until the approach is so close that a physical collision takes place. This is quantified by the Safronov number $\Theta = \frac{1}{2}\left(\frac{v_\mathrm{esc}}{v_\mathrm{orb}}\right)^2$ which compares the escape velocity $v_\mathrm{esc}$ from the planetary surface to the Keplerian velocity $v_\mathrm{orb}$ of the planet in its orbit around the star. For Jupiter-mass planets at 1\,au or beyond, the escape velocity is $\approx60\mathrm{\,km\,s}^{-1}$ but the orbital velocity $\lesssim 30\mathrm{\,km\,s}^{-1}$. This means that strong scattering is possible and collision uncommon. The typical outcome of scattering among giant planets therefore is the ejection of one or more planets from the system, and the retention of the survivors on eccentric and inclined orbits that are well separated and stable over long timescales. The eccentricity distribution of giant exoplanets is consistent with around three quarters of systems of giant exoplanets undergoing such instabilities \citep{JuricTremaine08,Raymond+11}.

The process of planet--planet scattering is itself chaotic, and it is impossible to uniquely predict a final orbital eccentricity for a given initial configuration. Nevertheless, some trends hold statistically. In general in two-planet systems that undergo scattering, the higher the mass of the ejected planet, the higher the eccentricity of the survivor (see figure~4 of \citealt{Mustill+22}). A planet that ejects another of just half its mass will be left with an orbital eccentricity $>0.3$ more than half of the time, while a planet that ejects an equal-mass planet will nearly always be left with such an eccentricity. Hence, in this paper, we adopt systems of two equal-mass planets for being both simple and likely to result in a planet with the target eccentricity of $e>0.3$.

A second general result of scattering amongst equal-mass planets is that the inner planet ends up with roughy half the semi-major axis of that of the original innermost planet. This can be understood as a simple consequence of energy conservation: if two planets of the same mass begin with comparable semimajor axes, and one is ejected with a hyperbolic velocity at infinity with respect to the host star only slightly above zero (a typical outcome), this means that the escaping planet has almost zero potential and kinetic energy after scattering, and so the surviving planet has absorbed all of the (negative) energy of its ejected partner and therefore roughly halved its semimajor axis.

\subsubsection{Discs from destruction of satellites}

\label{sec:disc_calculations}

The orbital semimajor axes of satellite systems around planets are fundamentally constrained by two limits. The lower limit is given by the Roche radius for tidal disruption, where the differential gravitational (tidal) field across the satellite is strong enough to overcome its own binding forces. For large satellites where material strength is negligible, this limit is set by the satellite's self-gravity and is given by
\begin{equation}
    R_\mathrm{Roche} = \left(\frac{3\rho_\mathrm{p}}{\rho_\mathrm{s}}\right)^{1/3}R_\mathrm{p},\label{eq:RRoche}
\end{equation}
where $\rho_\mathrm{p}$ and $\rho_\mathrm{s}$ are the mean densities of the planet and the satellite, and $R_\mathrm{p}$ is the physical radius of the planet; small changes to the constant are possible depending on the satellite's density profile and rotation state. The outer limit is set by the Hill sphere of gravitational influence around the planet
\begin{equation}
    R_\mathrm{Hill} = a_\mathrm{p}\left(\frac{M_\mathrm{p}}{3M_\star}\right)^{1/3}.\label{eq:RHill}
\end{equation}
Prograde satellites are stable if their orbits are within $\approx0.5R_\mathrm{Hill}$ \citep{Nesvorny+03}. 

\begin{figure}
\includegraphics[width=0.5\textwidth]{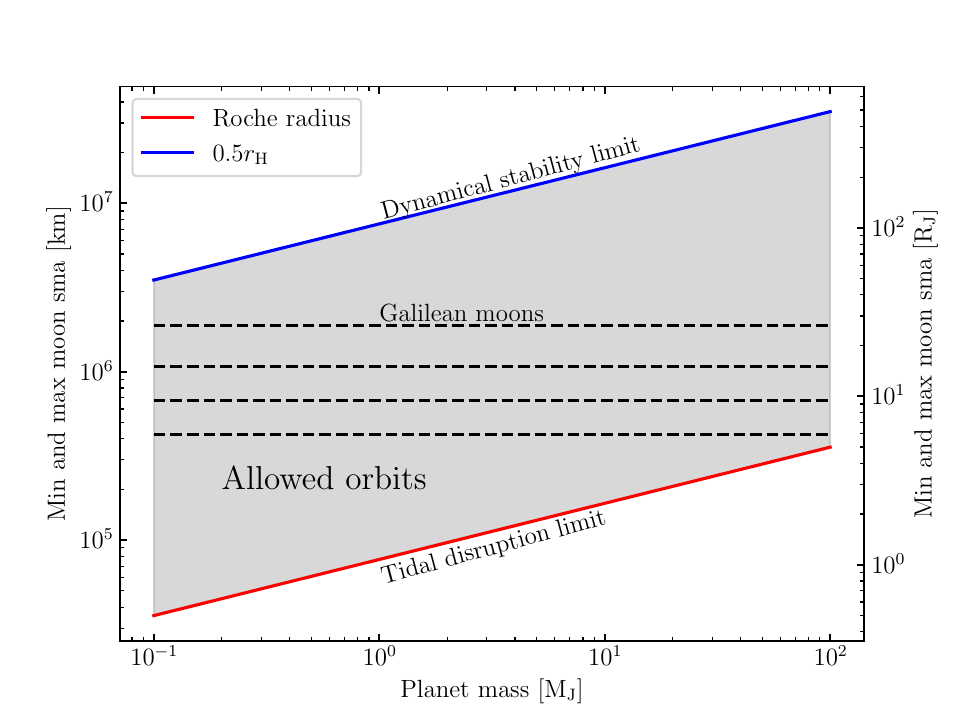}
\caption{Allowed semimajor axes of satellites around planets of a range of masses. The innermost orbit is determined by the Roche radius for tidal disruption, shown here for a satellite density of $3\mathrm{\,g\,cm}^{-3}$ (Equation~\ref{eq:RRoche}). The outer limit is set by the Hill radius around the planet, shown here for a planetary semimajor axis of $1.4$\,au and a stellar mass of $0.85\mathrm{\,M_\odot}$ (Equation~\ref{eq:RHill}). Horizontal lines mark the orbital radii of the Galilean moons around Jupiter.}
\label{fig:moon_radii}
\end{figure}

The range of allowed orbits for planets of a range of masses are shown in Figure~\ref{fig:moon_radii}. Here we took $\rho_\mathrm{s}=3\mathrm{\,g\,cm}^{-3}$ for the satellite density in the calculation of the Roche limit, and $a_\mathrm{p}=1.4$\,au for the planet's semimajor axis before scattering in the calculation of the Hill radius. The allowed orbital semimajor axes of the satellites span nearly two orders of magnitude for these parameters. The orbital radii of Jupiter's Galilean moons are also marked, and comfortably lie within the allowed region except for very massive ``planets'' at $M_\mathrm{p}\gtrsim100\mathrm{\,M_J}$ where Io would risk tidal disruption, and low-mass giant planets at $M_\mathrm{p}\lesssim0.1\mathrm{\,M_J}$ where the Solar perturbations to the orbits may trigger dynamical instability.

\begin{figure*}
    \centering
    \includegraphics[width=\textwidth]{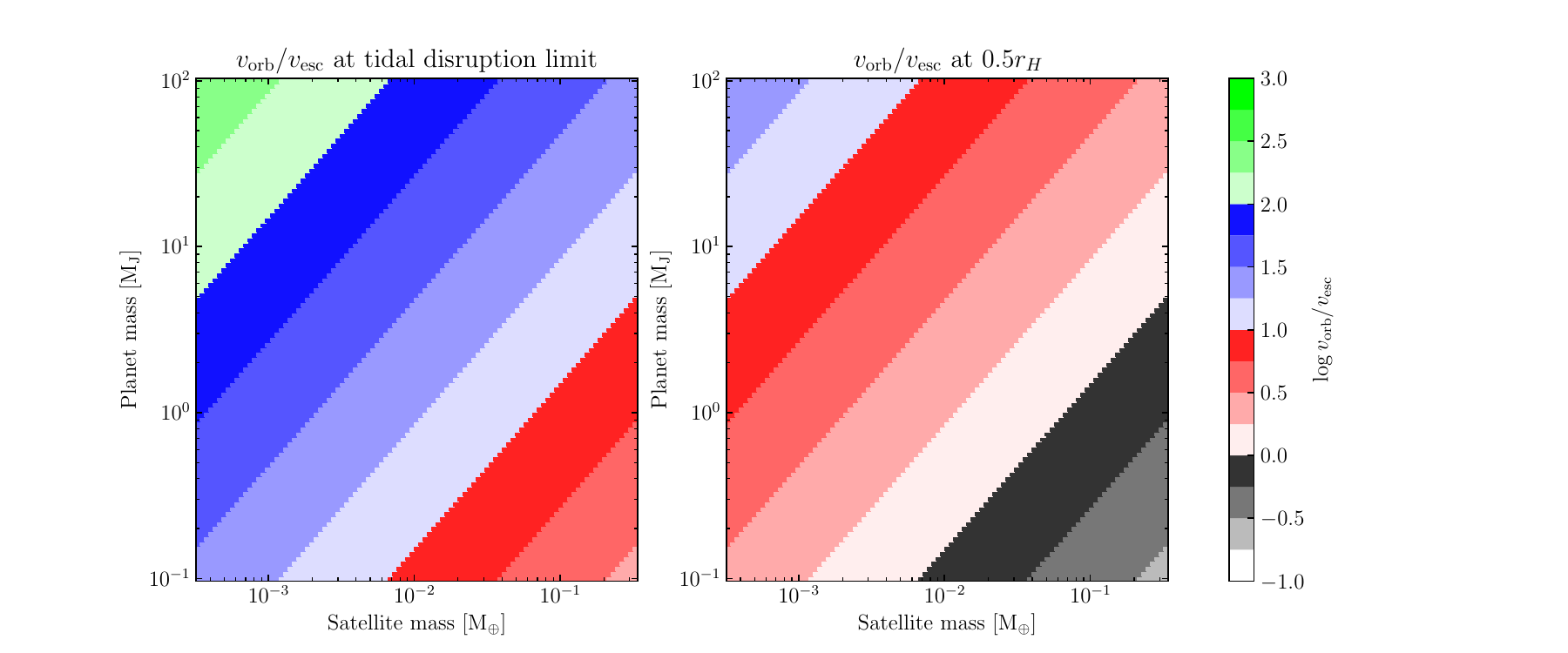}
    \caption{The ratio of a satellite's orbital velocity around its planet to its surface escape velocity, for satellites orbiting at the Roche radius (left) and at the outer stability limit at $0.5r_\mathrm{H}$ (right). We take a satellite density of $\rho_\mathrm{s}=3\mathrm{\,g\,cm}^{-3}$, a planetary semimajor axis of $1.4$\,au, and a stellar mass of $0.85\mathrm{\,M}_\odot$.}
    \label{fig:vratio}
\end{figure*}

The close encounters between giant planets during scattering will have an effect on their moon systems, conceptually illustrated in Figure~\ref{fig:cartoon}. These effects can include the capture of a moon of one planet by another planet (Figure~\ref{fig:cartoon}.I), or the excitation of orbital eccentricities and inclinations during the encounter. Directly or indirectly, this means that the orbits of the moons can begin to intersect, and the moons experience close encounters (Figure~\ref{fig:cartoon}.II). In contrast to the situation described above for giant planet scattering, the moons will not undergo strong scattering as a result of their mutual gravitational interactions: their escape velocities are low ($2-3\mathrm{\,km\,s}^{-1}$ for the Galilean moons) but their orbital velocities (around the host planet) high ($8-17\mathrm{\,km\,s}^{-1}$), meaning that an encounter close enough to significantly change orbital elements cannot occur without a physical collision. We map out the ratio of a satellite's orbital to escape velocity, as a function of satellite mass and planetary mass, in Figure~\ref{fig:vratio}. Except for high-mass moons orbiting low-mass gas giants close to the outermost stable orbit at $\approx0.5R_\mathrm{Hill}$, the orbital velocity is considerably in excess of the surface escape velocity, meaning collisions are overwhelmingly favoured over strong gravitational scattering. In addition, because these physical collisions can occur at velocities significantly in excess of the escape velocity from the satellite surface, the collisions will frequently be "supercatastrophic" and completely pulverise the moons, generating copious debris \citep[e.g.,][]{LeinhardtStewart12}. The outcome of instability among the satellite systems, then, will primarily be satellite--satellite collisions; and the outcome of such a collision will be a large debris cloud that will undergo further collisional grinding as it settles into a circumplanetary disc (Figure~\ref{fig:cartoon}.IIIA).

A second route to debris production in the moon systems is if one of the moons is perturbed onto an orbit that crosses the Roche limit for tidal disruption around the planet (Figure~\ref{fig:cartoon}.IIIB).
Simulations of asteroids disrupted by stars in this manner show that, even if large fragments remain after one close passage, disruption continues over subsequent orbits so that quickly the body is reduced to fragments with enough internal strength to resist further fragmentation \citep{Li+21}. In the absence of additional forces, the debris forms a collisionless ring around the disrupting body \citep{Veras+14}. Differential orbital evolution of the fragments, in our case arising from planetary oblateness, the star, and any other surviving satellites, will result in collisions between debris particles and therefore the reduction of the remnants of the satellite to dust as the debris trail settles into a disc, similar to the case of a satellite--satellite collision.

Finally, we can estimate from the size of the observed transiting structures the amount of mass that is required to occult the star, and hence the minimum initial mass of the disrupted satellite(s). The surface density of Saturn's optically-thick B ring is of the order $100\mathrm{\,g\,cm}^{-2}$ \citep{HedmanNicholson16}. If we assume that the disc of EPIC~2202b has the same surface density, and multiply by the area of a circle $1.1\mathrm{\,R}_\odot$ in radius, we obtain a minimum mass for the disc of $2\times10^{21}$\,kg, corresponding to a body of roughly $500$\,km in radius, and comparable in size to Ceres. This mass is around $2.7\%$ of the mass of Earth's Moon, and just $0.5\%$ of the total mass of Jupiter's Galilean satellites\footnote{Values taken from \url{https://ssd.jpl.nasa.gov/sats/phys_par/}}. Disruption of only a mid-sized satellite, therefore, suffices to produce enough debris to form a significant disc; alternatively, disruption of large satellites would over-produce debris, allowing for inefficient use of the material owing to a higher surface density, or for loss of material as a result of evolution of the disc.

\section{Numerical simulations}

\label{sec:method}

\subsection{Preliminary long-term simulations}

\label{sec:longterm}

We first performed a set of long-term, planet-only simulations in order to gain some statistics on the number of close encounters between planets during scattering, as well as the distances of closest approach. We ran 100 simulations with the \textsc{Mercury} code \citep{Chambers99}, each of two planets of $1\mathrm{M_J}$ orbiting a $0.85\,M_\odot$ star \citep[the mass of EPIC~2202;][]{vanderKamp+22}. The innermost planet was located at 2\,au so that energy conservation would cause it to move inwards to around 1\,au after ejection of the second planet, and the outer planet was placed randomly uniformly from 0 to 3.6 mutual Hill radii in order to ensure that scattering began quickly. Simulations were run with the RADAU integrator \citep{Everhart85} with an accuracy parameter of $10^{-11}$. Simulations ended when bodies were removed by collision or ejection beyond 50\,000\,au, or after 100\,Myr had elapsed. Close encounter distances between planets within 2 Hill radii were logged during the simulation.

\subsection{Main short-term simulations}

We subsequently ran our main simulations, including both planets and satellites, with the \textsc{Rebound} package \citep{rebound} using the high-accuracy IAS15 integrator \citep{reboundias15}, which is an improved version of RADAU with better accuracy and error handling. We adopt this in order to accurately resolve the short orbital periods of the satellites around their host planets, which also necessitate a much shorter time-step for simulating satellite systems than systems of only planets. The phase of evolution prior to planet--planet scattering may be long, and planets are not guaranteed to undergo scattering within the finite duration of a numerical integration. Therefore, for computational efficiency, we adopted the following approach \citep[similar to][]{RabagoSteffen19}:
\begin{itemize}
\item We run two-planet scattering simulations of two $2\mathrm{\,M_J}$ planets orbiting a $0.85\mathrm{\,M_\odot}$ star. The innermost planet is placed at twice the semimajor axis of the observed EPIC~2202 disc (with a period of $290$ days). The outermost planet is placed at between $1$ and $5$ single-planet Hill radii (Eq~\ref{eq:RHill}) from the inner. Eccentricities are uniformly drawn between $0$ and $0.1$, inclinations between $0$ and $3$ degrees, and other orbital angles between $0$ and $360$ degrees.
\item These simulations are terminated when the planets undergo a close encounter within one Hill radius, which we define here as
  \begin{equation}
    r_\mathrm{Hill} = d\left(\frac{m_\mathrm{pl}}{3M_\star}\right)^{1/3},
  \end{equation}
  where $d$ is the planet's instantaneous distance to the star and $m_\mathrm{pl}$ and $M_\star$ are the planetary and stellar masses. The \texttt{heartbeat} function in \textsc{Rebound} is used to track the distance between planets. If no close encounter occurs within 1\,Myr, the simulation is terminated.
\item We take the system at this point, and integrate backwards in time for one year.
\item At this point, we insert clones of the Galilean satellite system around each planet, with masses and orbital elements taken from JPL Horizons\footnote{\cite{Giorgini+96}; \url{https://ssd.jpl.nasa.gov/horizons/}}. Note that our planets are twice Jupiter's mass while we do not rescale the masses of the satellites; if satellite mass correlates with planetary mass than our simulations will underestimate the extent of debris production. Each satellite system is given a random re-orientation of its longitude of ascending node.
\item We integrate this 11-body system forwards for
10 kyr. We record collisions between any pair of bodies, as well as ejections of any body beyond 50\,000\,au (practically unattainable given the integration duration). When satellites collide we  merge the bodies and continue to track the orbit of the merger product, while noting that debris will have been produced during the collision.
\item The Roche radius for tidal disruption of the satellites lies outside the physical radius of the planet for our chosen densities. Therefore, for satellite--planet collisions, we actually remove the satellite at the point at which it crosses the Roche limit, while also noting whether it is indeed on a collision course with the planet or if its pericentre lies outside the planet's physical radius. Satellites on a collision course with the planet are not recorded as producing debris trails.
\item We record the position and velocity of merger products between satellites, as well as of a satellite removed after crossing the Roche radius of a planet but that will not collide with it, as these will yield the orbital elements of the barycentre of debris trails that arise from collision or tidal disruption of the satellites.
\end{itemize}

\section{Numerical results}

\label{sec:results}

\subsection{Preliminary simulations}

\begin{figure}
    \centering
    \includegraphics[width=0.5\textwidth]{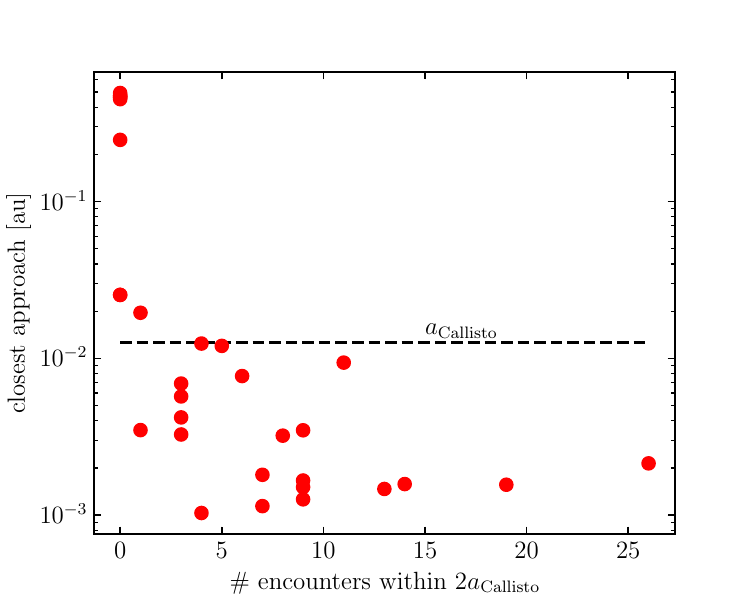}
    \caption{Number of extremely close approaches (within twice the semi-major axis of Callisto), and distance of the closest approach, in the set of preliminary 2-planet systems where one planet was ejected. The horizontal line marks the semi-major axis of Callisto.}
    \label{fig:CE}
\end{figure}

Planet--planet scattering can result in the collision of one planet with the other or with the star; in the ejection of a planet; or in the survival of both planets within the integration duration, with or without scattering having commenced. Of the 100 two-planet systems in our preliminary simulations, 31 systems lost a planet to ejection, while a further 37 experienced a planet--planet collision. Of the ejected planets, 7 experienced only weak encounters, and had no very close encounter $<0.1$\,au ($\sim8a_\mathrm{Callisto}$, where $a_\mathrm{Callisto}$ is the semimajor axis of Callisto, the most distant of the Galilean moons). These would not be expected to undergo scattering or exchange of satellites. The remainder had encounters $\lesssim2a_\mathrm{Callisto}$, often multiple times (Figure~\ref{fig:CE}). There are two consequences to note: first, planets that have ejected a second planet typically have had at least one encounter where satellite systems can be exchanged or destabilised; secondly, subsequent encounters could strip away some of the resultant debris, as discussed in \S\ref{sec:discussion}.

\subsection{Main simulations}

In the main simulations, 90 out of 100 systems experienced a close encounter of less than 1 Hill radius between planets, and satellite systems were inserted around each planet in these systems as described in \S\ref{sec:method}.

The fate of satellites is more complex than that of the planets. Each satellite can suffer one of the following fates:
\begin{enumerate}
    \item Remaining bound to the original host planet.
    \item Exchange to an orbit around the other planet, and survival throughout the integration.
    \item Collision with another satellite.
    \item Tidal disruption by one of the planets.
    \item Becoming unbound from any planet, where it may then remain on an orbit bound to the star, become unbound from the whole system, or return to impact or be tidally disrupted by another body.
\end{enumerate}
A key question is which outcomes lead to the production of debris that can settle into a circumplanetary disc. As argued in \S\ref{sec:disc_calculations}, outcome (iii) will always produce debris owing to the large collision velocities. Outcome (iv) will produce debris as the satellite crosses the Roche limit; this debris may immediately collide with the planet if the satellite orbit had a sufficiently small pericentre, it may be unbound from the planet if the satellite's velocity was very large, or it will otherwise remain in a bound orbit around the planet. Outcome (v) will produce bound debris if a collision with a moon or a favourable tidal disruption event around the planet occurs. We refer, at this point, to the production of a "debris trail" for each instance of debris arising from a collision or tidal disruption. These will, on a longer timescale, settle into a disc configuration. 

After insertion of the satellites, simulations were run for 10\,kyr. Eight systems ran extremely slowly and did not reach this point within a reasonable time-frame. These all had two surviving planets. Although these simulations had not run as long as the remainder, they had mostly experienced some debris production, with only three systems having had no debris production, and the remaining five having between two and five bound debris trails per system. We include these systems in the results below, although they will result in a slight underestimation of the production of debris. 

\begin{figure}
    \centering
    \includegraphics[width=0.5\textwidth]{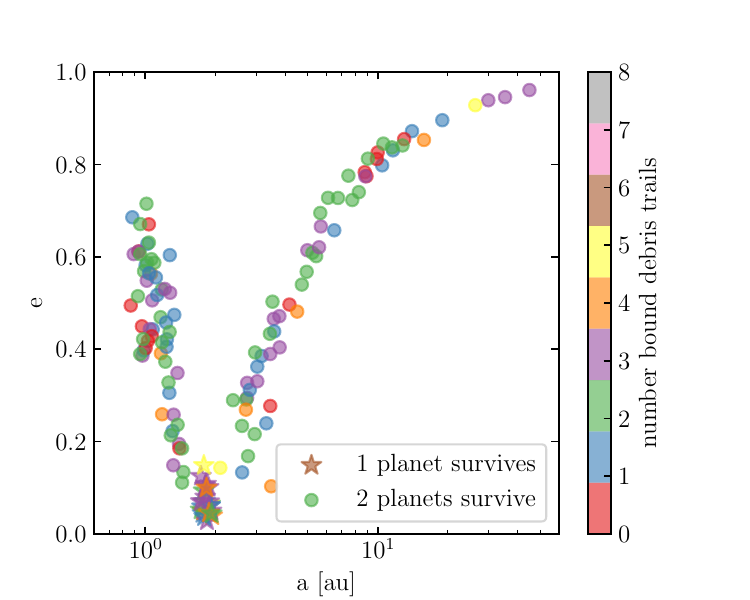}
    \caption{Orbital elements of planets in the 90 simulations where moons were inserted, at the end of the simulations. Stars mark single-planet systems where scattering has resulted in a planet--planet merger. Circles mark systems with two survivors; all of these are in the process of scattering and will likely resolve as either collisions or ejections. The colour of the points shows the number of debris trails bound to each planet.}
    \label{fig:a-e}
\end{figure}

At the end of the simulations, 30 out of 90 systems had reduced to a single planet, all through planet--planet collisions. 60 remained with two planets, all in the process of scattering. The orbital elements of these planets are shown in Figure~\ref{fig:a-e}. The eccentricities of the planets that have collided are all low to moderate ($e<0.2$), too low to be consistent with the host planets of the observed circumplanetary discs. Although none of the systems has yet ejected its outer planet, the innermost planets have already attained high eccentricities, with most being above the observational lower limit $e\gtrsim0.3$. Unfortunately, as the orbital period of the outer planet grows while it is scattered to ever-higher semimajor axes, it is not computationally feasible to follow the evolution of the satellites until the planetary system dynamics is resolved, and we have to settle on analysing this intermediate state.

\begin{figure*}
    \includegraphics[width=\textwidth]{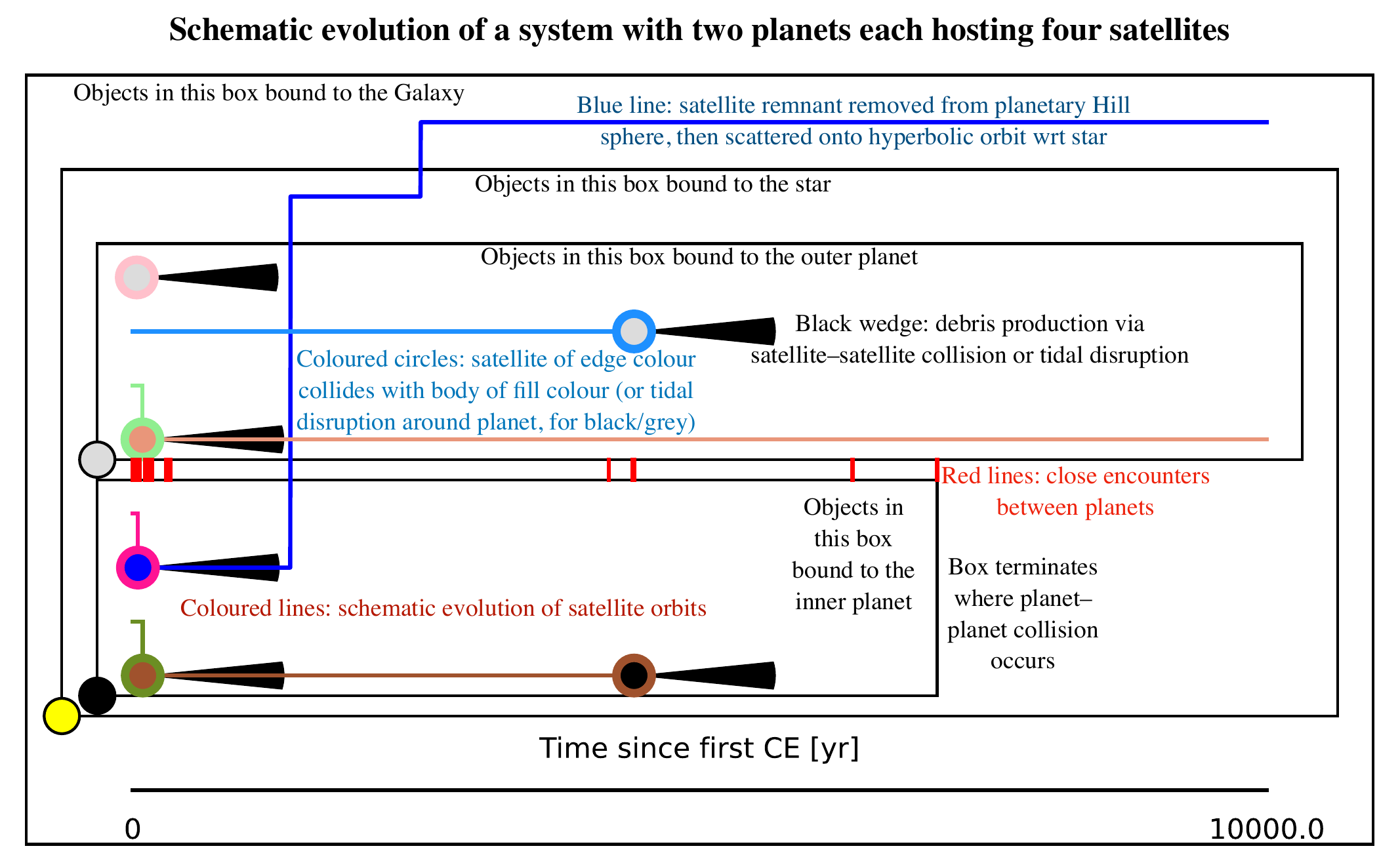}
    \caption{Topological illustration of the interactions between planets and satellites as a function of time in an example simulation. Nested boxes indicate whether objects are bound to the Galaxy, the star or a planet. The yellow circle indicates the star, and the black and grey circles the inner and outer planets. Coloured lines show, topologically, the orbits of satellites, ordered according to their initial locations around each planet (bottom being closest). Coloured circles and attached wedges mark the collision or tidal disruption of satellites and production of debris. Short red lines between the planetary boxes mark close encounters, and the lower planetary box is terminated when the two planets collide.}
    \label{fig:example}
\end{figure*}

Even within the relatively short 10\,kyr integration, the repeated interactions of the planets and their satellite systems can be extremely complex. One such example is shown in Fig~\ref{fig:example}, which shows the time evolution of the topological configuration of the system. Initially, four satellites are bound to each planet. A succession of early close encounters results in three satellite--satellite collisions and one tidal disruption. Of the three merger products, one subsequently escapes its planet's Hill sphere and later is scattered onto an orbit unbound from its host star, one is tidally disrupted after about 4\,000\,yr, and the third survives to the end of the integration. The final satellite is also tidally disrupted during a close encounter after around 4\,000\,yr. Finally, the two planets collide after around 7\,000\,yr. In all, there are thus five or six debris-producing events (depending on whether a sizeable remnant survives from the olive--brown collision), and the planet may have a large satellite surviving along with its disc if a large fragment comes from the green--salmon collision.

While we have chosen, for illustrative purposes, a particularly rich example system, production of debris through collision or tidal disruption of satellites is nearly ubiquitous in our simulations (further examples of evolution can be seen in Figure~\ref{fig:appendix}). 
In total, there are 89 collisions between satellites, and 361 tidal disruption events, making the latter the more common route to forming debris. The vast majority of these occur when the satellites are bound to a planet [cases (iii) and (iv) above]; the case where a liberated satellite is disrupted during  close encounter with a planet occurred only thrice, in two cases where the debris stream was on a collision course with the planet, and in one case where the debris stream was unbound.

\begin{figure}
    \includegraphics[width=0.5\textwidth]{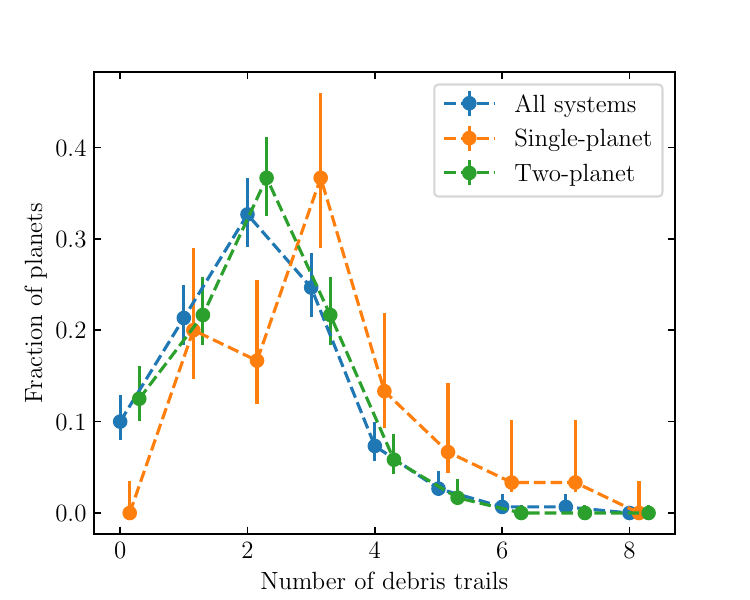}
    \caption{Fraction of planets with specified number of debris trails after the simulations, showing planets in single-planet systems, two-planet systems, and the combined sample (offset for clarity). Points show the occurrence frequency in the simulations, and error bars the 68\% Bayesian confidence interval \citep[][Chapter~6]{Jaynes03}.}
    \label{fig:fractions}
\end{figure}

Surviving planets in our simulations host between zero and seven\footnote{The maximum possible would be eight, one from each satellite.} such debris trails, inclusive; in all of the single-planet systems, the planet hosts at least one debris trail. The number of bound trails per planet is illustrated by the colour scale in Figure~\ref{fig:a-e}: the innermost planets of the systems still undergoing scattering host between zero and four debris trails. In Table~\ref{tab:ndebris} and Figure~\ref{fig:fractions} we give the number and fraction of debris trails hosted by each planet for all simulations: both in systems that reduce to a single planet, in systems that retain two planets, and in the combined sample. Only 10\% of surviving planets host no debris trails, and eight of these were in the unfinished runs that had not yet reached 10\,kyr. Planets in single-planet systems host somewhat more debris trails than in two-planet systems, with a mean of 2.9 \emph{vs.~}1.9. We recall that the two-planet systems are still evolving dynamically, and there in longer simulations there would be the further chance of exchange of satellites and/or their destabilisation. Further internal dynamics within the disrupted satellite systems could also lead to more collisions between satellites, and possibly tidal disruptions, on a longer time-scale. Nevertheless, despite running for only up to 10\,kyr, the duration was sufficient to resolve the satellite dynamics in most systems. At the end of the simulations, only 19 systems still possessed two or more satellites, and in all but three cases these had already experienced some collisions or disruptions.

\begin{table*}
    \centering
    \begin{tabular}{l|ccccccccc}
         Number of bound debris streams & 0 & 1 & 2 & 3 & 4 & 5 & 6 & 7 & 8\\
         \hline 
         All planets ($n=150$) & 15 & 32 & 49 & 37 & 11 & 4 & 1 & 1 & 0\\
         Planets in 1-planet systems ($n=30$) & 0 & 6 & 5 & 11 & 4 & 2 & 1 & 1 & 0\\
         Planets in 2-planet systems ($n=120$) & 15 & 26 & 44 & 26 & 7 & 2 & 0 & 0 & 0\\
         \hline
         All planets ($n=150$) & 10.0\% & 21.3\% & 32.7\% & 24.7\% & 7.3\% & 2.7\% & 0.7\% & 0.7\% & 0.0\%\\
         Planets in 1-planet systems ($n=30$) & 0.0\% & 20.0\% & 16.7\% & 36.7\% & 13.3\% & 6.7\% & 3.3\% & 3.3\% & 0.0\%\\
         Planets in 2-planet systems ($n=120$) & 12.5\% & 21.7\% & 36.7\% & 21.7\% & 5.8\% & 1.7\% & 0.0\% & 0.0\% & 0.0\%
    \end{tabular}
    \caption{Upper: number of planets with given number of bound debris streams. Lower: fraction of planets with given number of bound debris streams.}
    \label{tab:ndebris}
\end{table*}

\begin{figure}
    \centering
    \includegraphics[width=0.5\textwidth]{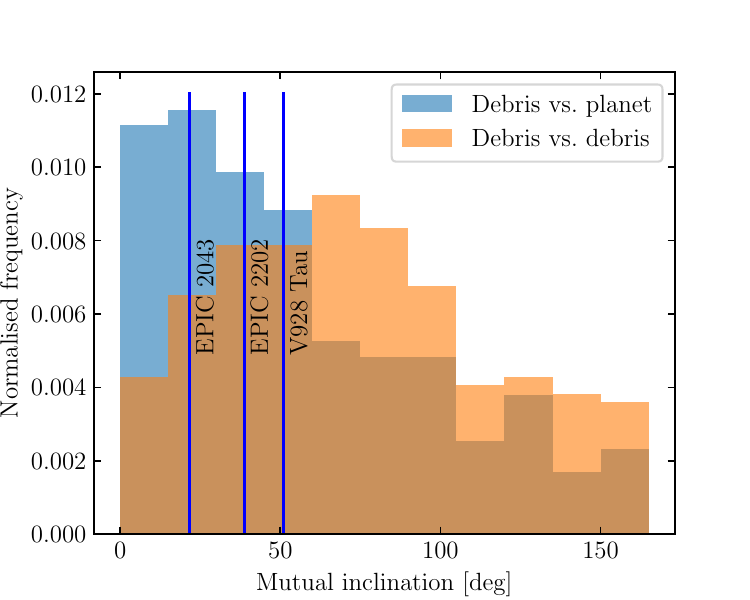}
    \caption{Inclinations of debris trails at the end of the simulations. Blue: inclination of the debris orbit around the planet relative to that of the host planet's orbit around the star. Measured inclinations of the three observed systems are shown as vertical blue lines. Orange: mutual inclination of each pair of debris trails that orbit the same planet.}
    \label{fig:inc}
\end{figure}

An edge-on disc surrounding a transiting planet would produce only a weak transit signature, and so it is of interest to study the mutual inclination between the disc and the planetary orbit around the star. During our scattering simulations, the orbital inclinations of the planets remain only slightly excited, lying within a few degrees of the reference plane. However, the debris streams are highly excited with respect to the planetary orbits, with mutual inclinations typically of tens of degrees and many being retrograde (Fig~\ref{fig:inc}, blue histogram). As the planetary orbit is seen almost edge-on in order for a transit to occur, this means that the resultant discs will also be observed at a similar inclination. We compare this to the corresponding inclinations of the three observed systems. We calculate the inclination $I$ between the disc normal and the orbital normal as
\begin{equation}
\cos I = \sin i \cos \phi,
\end{equation}
where $i$ is the observed inclination between the disc normal and the line of sight, and $\phi$ is the angle between the disc's projected major axis and the transit chord across the star, with values given by \cite{vanderKamp+22}. These shown as the vertical lines in Figure~\ref{fig:inc} and are in excellent agreement with the distribution predicted from our simulations.

We also show in Fig~\ref{fig:inc} as the orange histogram the inclinations between different debris streams orbiting the same planet. This distribution more closely approaches isotropy than does the distribution of the angle between the debris and planetary orbits. These high inclinations help justify our assumption that the satellite--satellite collisions will be highly destructive. They also imply that collisions between particles of different streams will occur at high velocity, preventing reaccretion of debris into satellites, and hastening collisional grinding to speed the formation of a dust disc.

\section{Discussion and conclusion}

\label{sec:discussion}

We have shown that the collision or tidal disruption of satellites is a ubiquitous outcome of planet--planet scattering. This explains many features of the large discs observed transiting EPIC~2202, EPIC~2043, and V928 Tau. The collision or disruption of satellites larger than Ceres would liberate enough mass to account for the size of these discs. The planet--planet scattering itself explains the observed eccentric circumstellar orbits of the bodies at the centre of the discs. In $N$-body integrations of planets with copies of the Galilean satellite system, we found that all but 10\% of planets undergoing scattering lost at least one satellite to collision or tidal disruption. Planetary systems that reduced to single-planet systems through planet--planet collision finished the simulations with an average of $2.9$ debris trails from lost satellites, while those whose scattering was unresolved by the end of the simulations possessed on average $1.9$ debris trails each. These debris trails often orbit their host planet at moderate to high inclination, again in agreement with the observed discs. In time, dissipation amongst the debris particles will cause their orbits to align with the Laplace plane, which approaches the planet's orbital plane at large planetocentric radii: maintaining a high inclination of the discs may prove a challenge. If satellites, or large collisional fragments, survive on inclined orbits, this would cause a strong misalignment in the disc. In any event, we expect the disc to be warped and not flat. Significant residuals in the fits of \cite{vanderKamp+22} may suggest that the real structures are indeed more complex than flat, featureless discs.

Nonetheless, our simulations are restricted in several ways. Computation times were long owing to the very small orbital periods of the satellites about the planets, and so we could not follow the planetary scattering all the way to ejection of a planet, which can take several Myr. As scattering proceeds, it is likely that further exchange, collision and disruption of satellites will occur during ongoing close encounters. The average of $1.9$ debris trails per planet in our unresolved simulations will therefore be an underestimate. On the other hand, close encounters will also result in the stripping of some material from the debris trails and its ejection into interplanetary space. Returning to the long-term simulations described in \S\ref{sec:longterm}, we find a median of four close encounters within $a_\mathrm{Callisto}$, and seven within $2a_\mathrm{Callisto}$, among the systems that lost a planet due to strong encounters leading to ejection. If we estimate that such a deep encounter would unbind roughly half of the debris particles \citep[\emph{e.g.,}][]{Li+19}, we could estimate a reduction in debris mass of one or two orders of magnitude. As the destruction of Galilean satellites over-produces the required debris by about two orders of magnitude (\S\ref{sec:disc_calculations}), this is not a significant problem.

A related issue concerns the longevity of the discs once planetary scattering is over. We can imagine these as analogous to more extensive versions of Saturn's ring system, age estimates for which range from comparatively young \citep[10–100\,Myr, \emph{e.g.,}][]{Iess+19} to the $4.6$\,Gyr age of the Solar System \citep[\emph{e.g.,}][]{Crida+19}. In light of the uncertainties in modelling the long-term evolution of such disc/ring systems, we do not pursue such modelling here. We do however note that two of the three observed disc hosts are very young: V928 Tau is a pre-MS star and candidate Taurus--Auriga member which would put its age at a few Myr \citep{vanDam+20}, while EPIC~2043 is a probably Upper Sco member at around 10\,Myr \citep{Rappaport+19}. These are considerably younger than Saturn's rings and it is likely that the observed discs could persist for the few Myr required. 

An issue related to the longevity for the discs is that of re-accretion of material, since the discs lie at least partially outside the Roche radius of their host planet. We estimate the prospects for re-accretion using the Toomre $Q$ parameter
\begin{equation}
Q = \frac{v\Omega}{\pi G \Sigma},
\end{equation}
where $v$ is the velocity dispersion, $\Omega$ the orbital frequency, $G$ the gravitational constant, and $\Sigma$ the surface density. Taking values $v=1$\,cm\,s$^{-1}$, $\Omega(M=2\mathrm{\,M_J},R=0.5\mathrm{\,R_\odot})=7.8\times10^{-5}$\,s$^{-1}$, and $\Sigma=200$\,g\,cm$^{-2}$,  yields $Q=3.7$, marginally stable. An alternative argument based on a comparison of Roche density and midplane density \citep{Beurle+10} yields $\rho_\mathrm{Roche}/\rho_\mathrm{midplane}=8.3$, again implying stability. On the other hand, with a smaller velocity dispersion of 1mm/s, the disc would be unstable and fragment. This may imply a slightly higher velocity dispersion than is seen in Saturn’s rings \citep[which may be as low as $\sim1$\,mm\,s$^{-1}$][]{GoldreichTremaine78}, which may be due to stirring by surviving satellites (original or fragments of the parent bodies). Alternatively, gas may be present to prevent the collapse of the solids into a thin sheet. This may arise from vapourised volatiles from the moons, or hint that the discs are actually remnants of the primordial circumplanetary disc in the protoplanetary era.

Another possible objection is that choosing the Galilean satellite system as a template may be unduly optimistic, as the other giant planets of the Solar System have fewer very massive moons. As discussed in \S\ref{sec:intro}, there are currently no meaningful observational constraints on the prevalence of extrasolar satellite systems, except for those belonging to very close-in planets. We can make a simple estimate of what would happen in systems containing fewer satellites by scaling down our results for the Galilean systems. We can asume that the frequency of satellite--satellite collisions will scale roughly with $N^2$ where $N$ is the number of satellites, while the frequency of tidal disruptions may scale by $N$ (if primarily driven by planetary encounters) or by $N^2$ (if primarily driven by satellite--satellite dynamical interactions). In our systems, we had 89 satellite--satellite collisions and 361 tidal disruption events. Dropping from 4 to 1 satellites per planet would therefore result in $\sim6$ collisions, and $\sim 23$ or $\sim 90$ tidal disruptions for $N^2$ and $N$ scalings respectively. This still gives an average of between $0.3$ and 1 debris-producing events per system. Thus, debris production should remain common, unless the planets are totally devoid of moderately-sized or larger satellites. If we take instead the Uranian satellite system as a template, we find only four satellites marginally larger than our minimum required radius of 500\,km, meaning we would require an extremely high efficiency of production and retention of debris: such low-mass satellite systems can probably be discounted as progenitors of large circumplanetary discs. Whether the Galilean or Uranian systems are more likely templates is at present observationally unconstrained, but if satellite mass scales roughly with planetary mass we can expect the Galilean system to better serve as a template given the large mass of the planet. The Galilean system has a peculiar resonant spacing which may not obtain in all satellite systems, but as most of the debris is produced by tidal disruption driven by planet--satellite dynamics during planet--planet scattering, the exact configuration of the satellites should not matter too much.

Finally, we note several possible alternatives to capturing eccentric satellites that could undergo collision or tidal disruption. First is the capture of one component of a binary asteroid into orbit around a planet \citep{AgnorHamilton06,Vokroulicky+08,Philpott+10}. This would, however, not excite the eccentricity of the planet itself. Nor would the disc being a remnant circumplanetary disc from the era of planet formation. A final alternative would be the capture of large asteroids as irregular satellites during planet--planet scattering \citep{Nesvorny+07,Nesvorny+14Jupiter}. This can explain the irregular satellites of the gas giants of the Solar System, which are less massive than is required to generate large circumplanetary discs (we show in \S\ref{sec:disc_calculations} that roughly Ceres-sized or larger satellites are needed). Both of the asteroid capture mechanisms therefore suffer from a low likelihood of capturing extremely large asteroids, and we have therefore focused on perturbations and exchange of satellites in this Paper.

\section*{Acknowledgements}

The authors thank for anonymous referee for 
comments that helped to improve the paper.
AJM acknowledges support from Career grants 120/19C and 2023-00146 from the 
Swedish National Space Agency. 
This research made use of Astropy,\footnote{\url{http://www.astropy.org}} 
a community-developed core Python package for 
Astronomy \citep{astropy:2013, astropy:2018}. 
This research made use of NumPy \citep{2020NumPy-Array}, 
%SciPy \citep{2020SciPy-NMeth}
and MatPlotLib \citep{2007CSE.....9...90H}.
Simulations in this paper made use of the REBOUND N-body code \citep{rebound}.
The simulations were integrated using IAS15, a 15th order Gauss-Radau integrator
\citep{reboundias15}. The SimulationArchive format was used to store fully
reproducible simulation data \citep{reboundsa}. 
This research has made use of NASA's Astrophysics Data System Bibliographic Services and the SIMBAD database, operated at CDS, Strasbourg, France. 

\section*{Data availability statement}

Scripts to reproduce the \textsc{Rebound} simulations and analysis are available at \url{https://github.com/AJMustill/moons}.

\bibliographystyle{mnras}
\bibliography{cpds}

\appendix

\section{Further examples of evolution}

\begin{figure*}
    \includegraphics[width=0.3\textwidth]{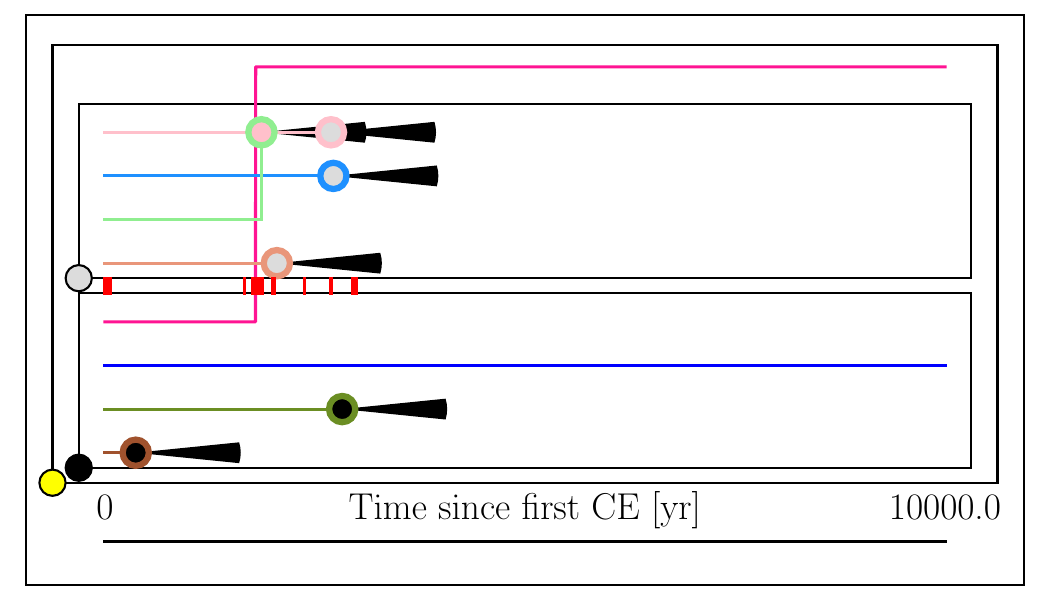}
    \includegraphics[width=0.3\textwidth]{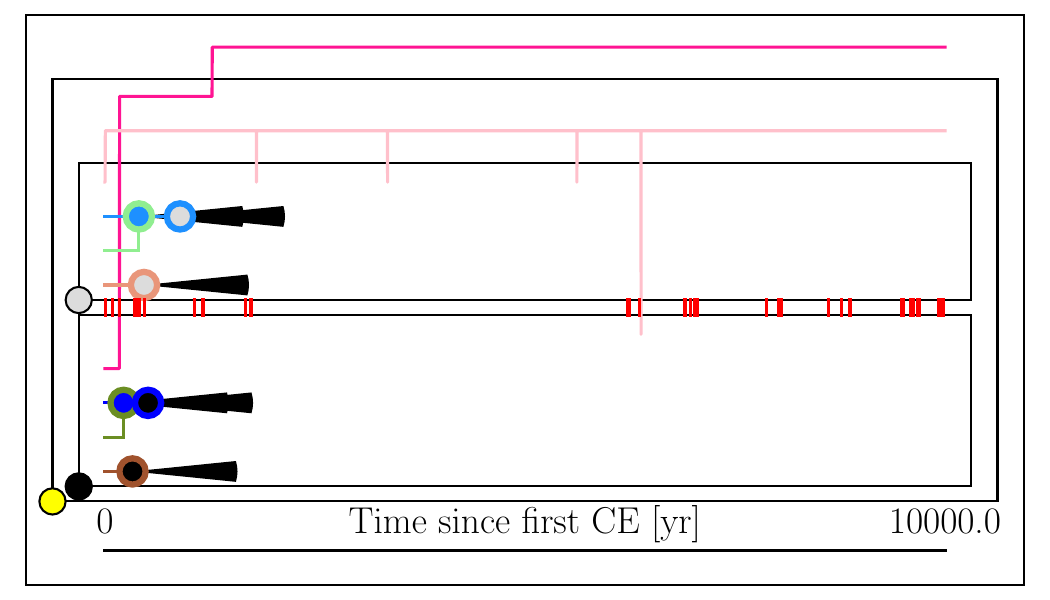}
    \includegraphics[width=0.3\textwidth]{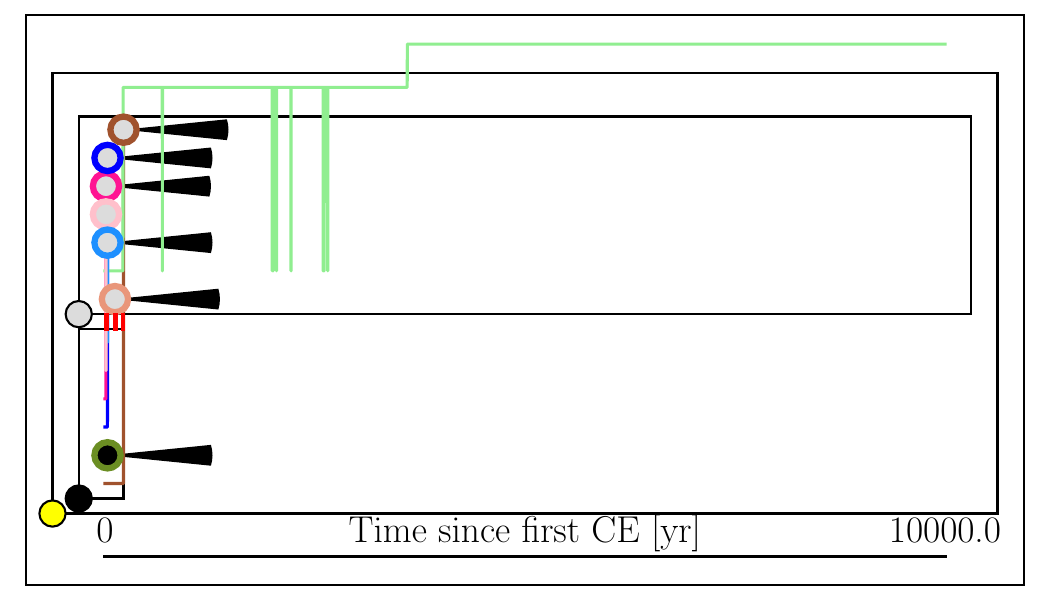}
    \includegraphics[width=0.3\textwidth]{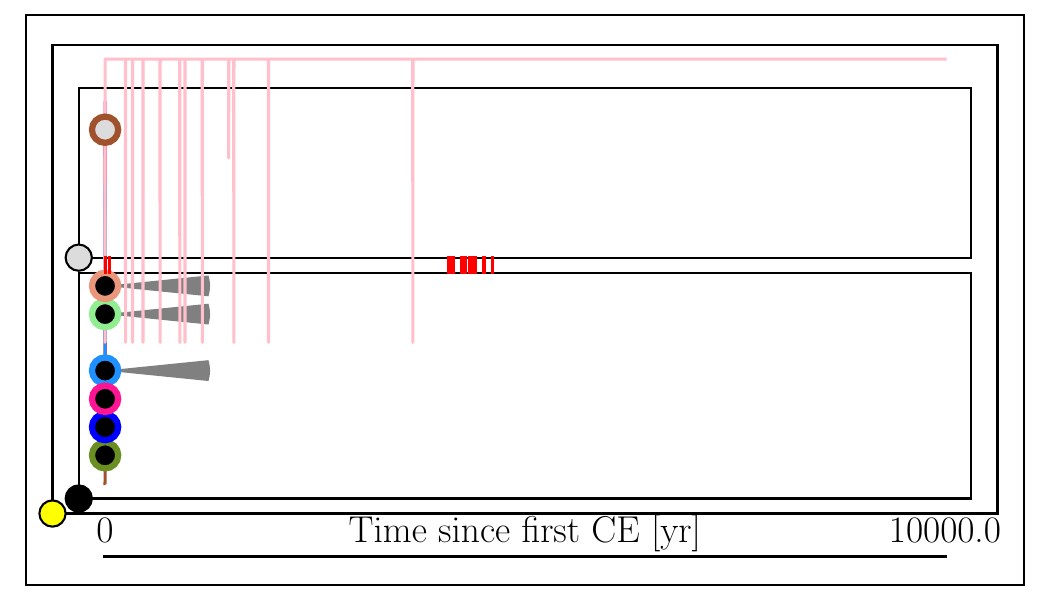}
    \includegraphics[width=0.3\textwidth]{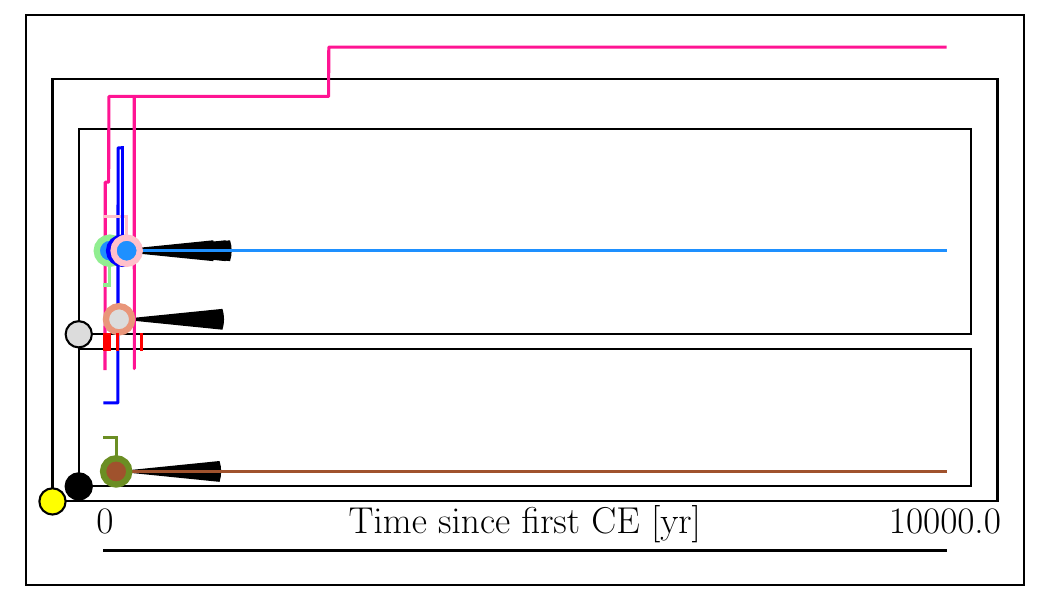}
    \includegraphics[width=0.3\textwidth]{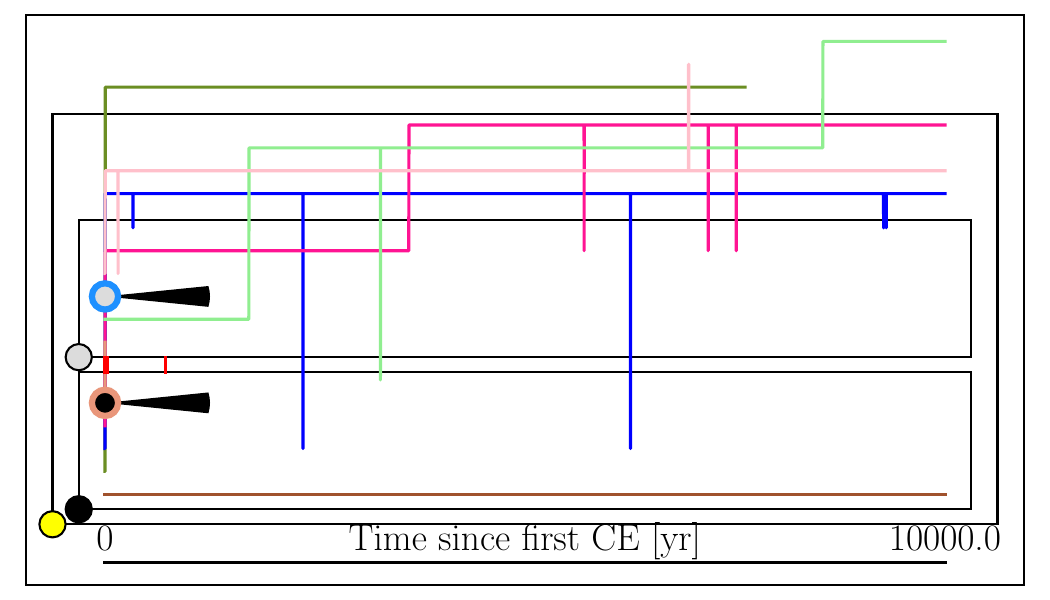}
    \includegraphics[width=0.3\textwidth]{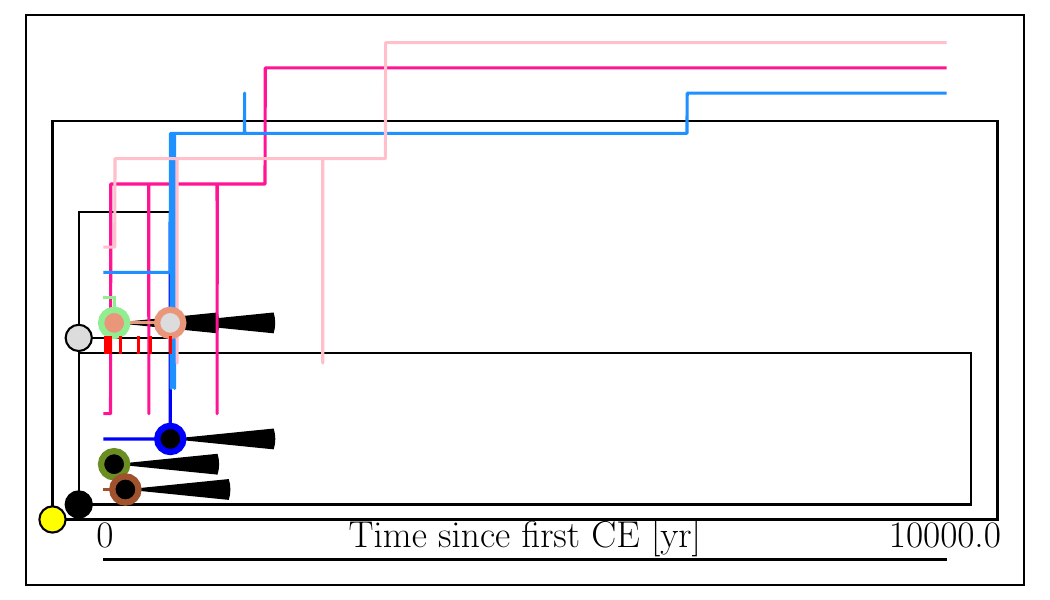}
    \includegraphics[width=0.3\textwidth]{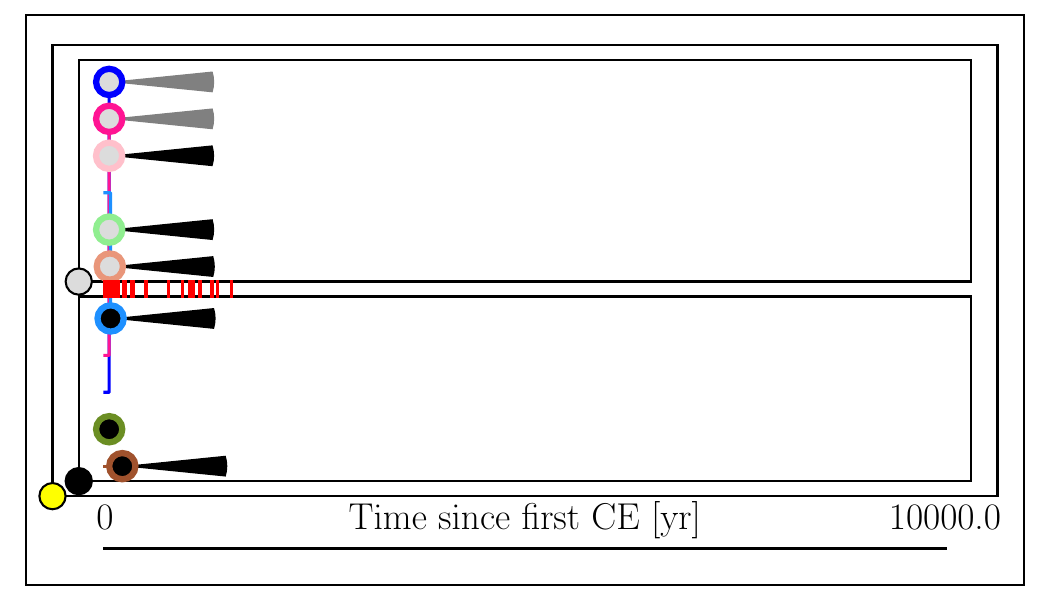}
    \includegraphics[width=0.3\textwidth]{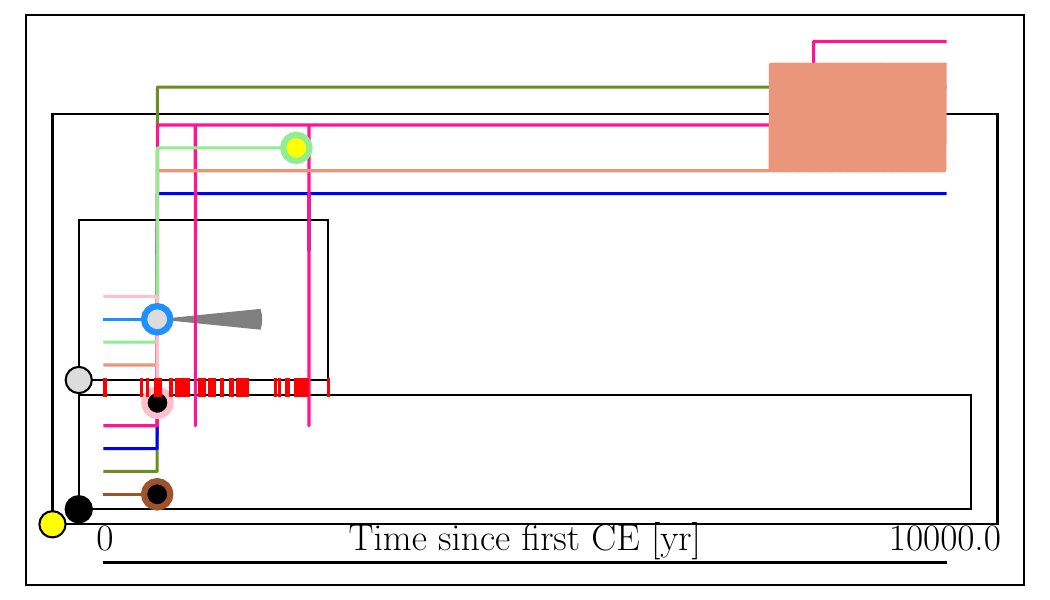}
    \includegraphics[width=0.3\textwidth]{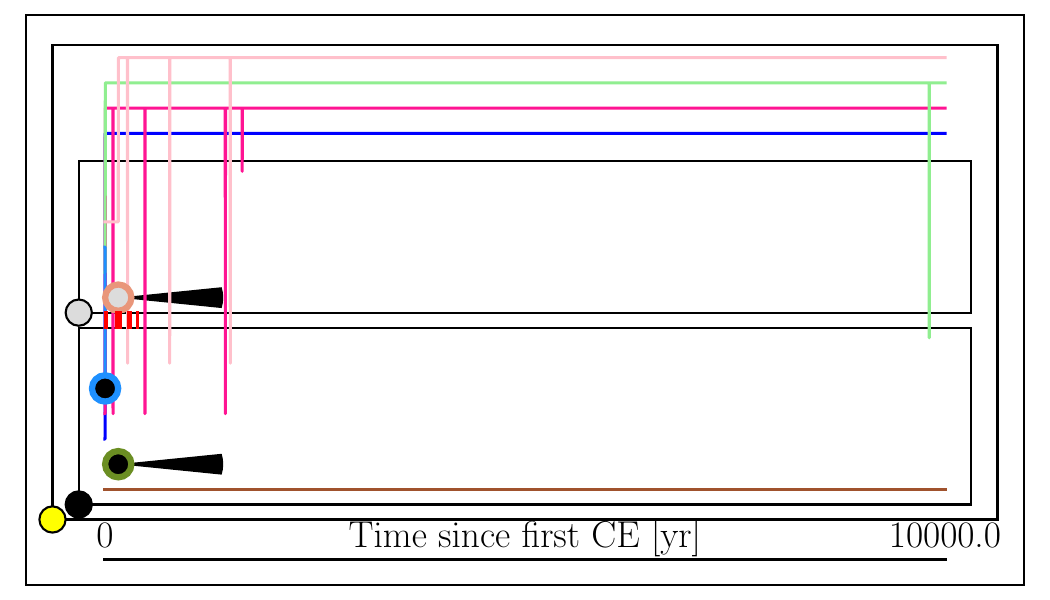}
    \includegraphics[width=0.3\textwidth]{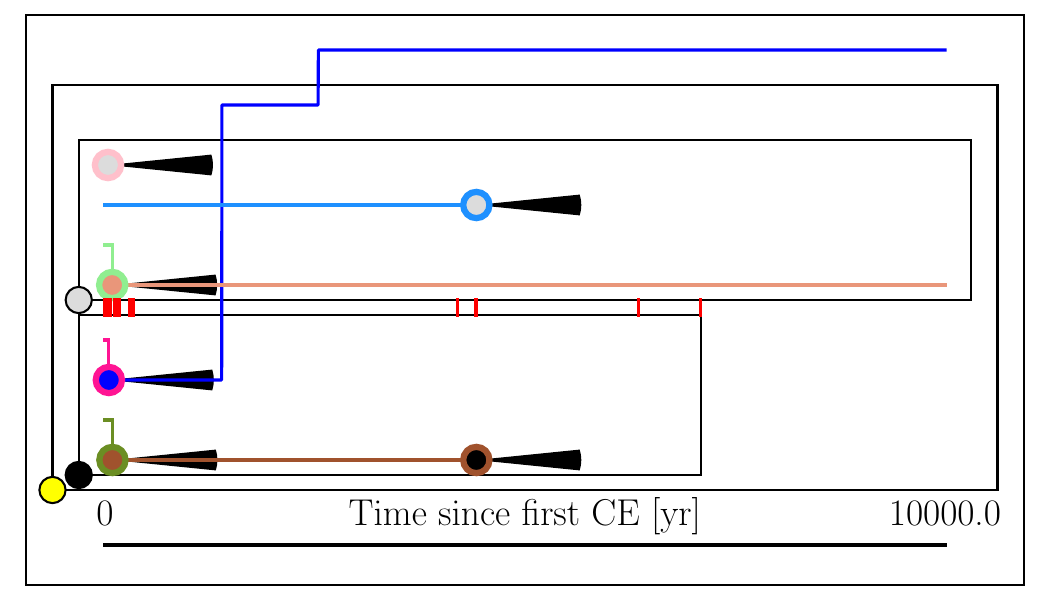}
    \includegraphics[width=0.3\textwidth]{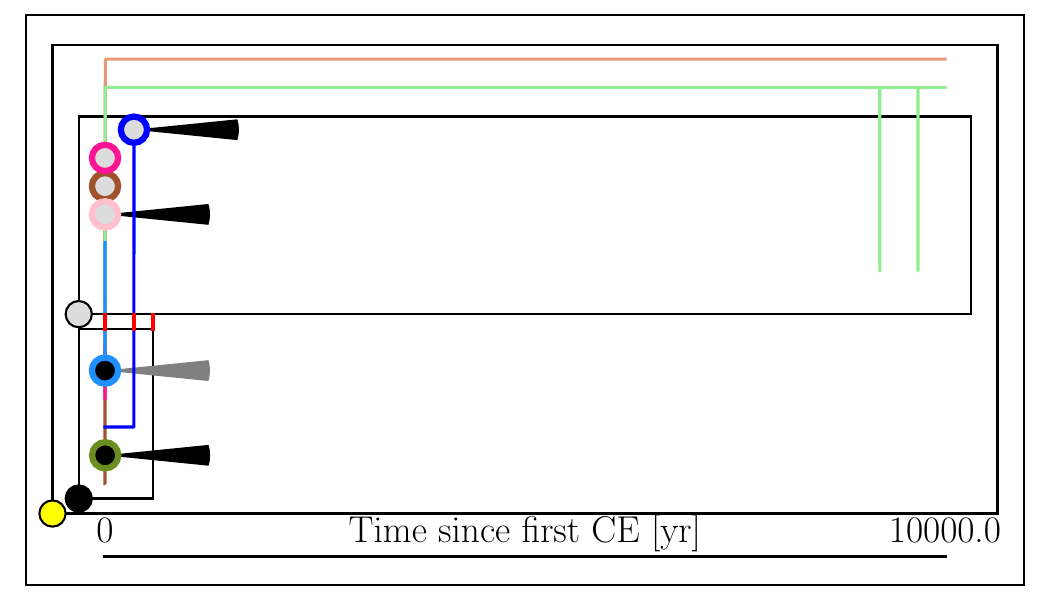}
    \includegraphics[width=0.3\textwidth]{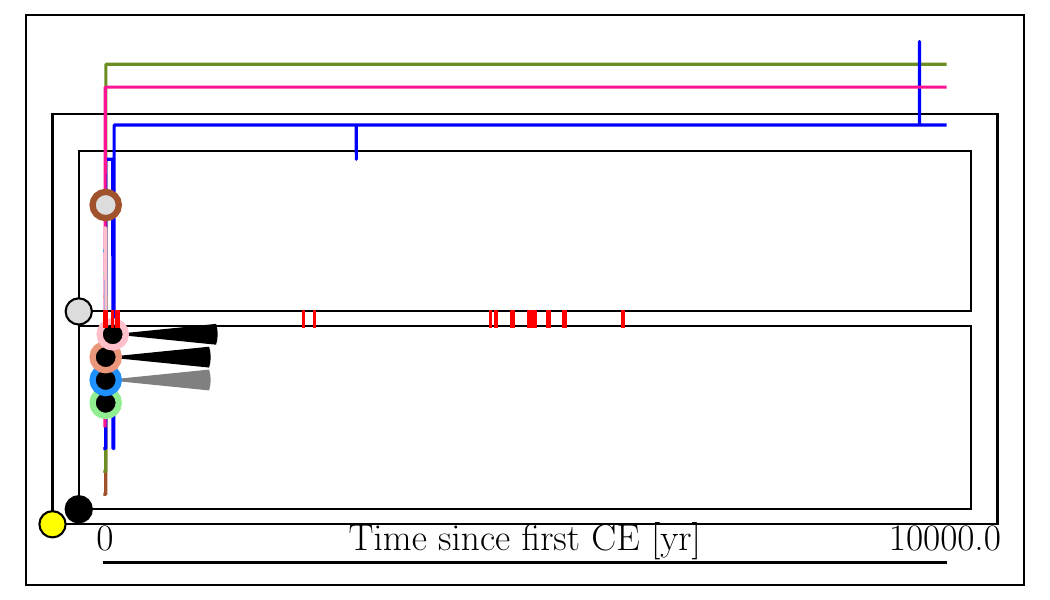}
    \includegraphics[width=0.3\textwidth]{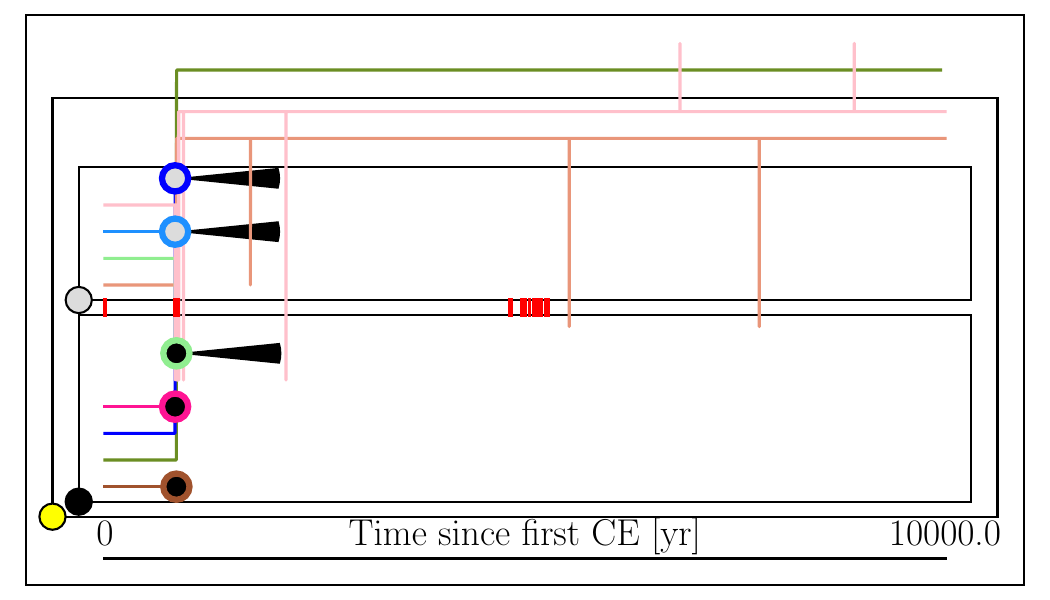}
    \includegraphics[width=0.3\textwidth]{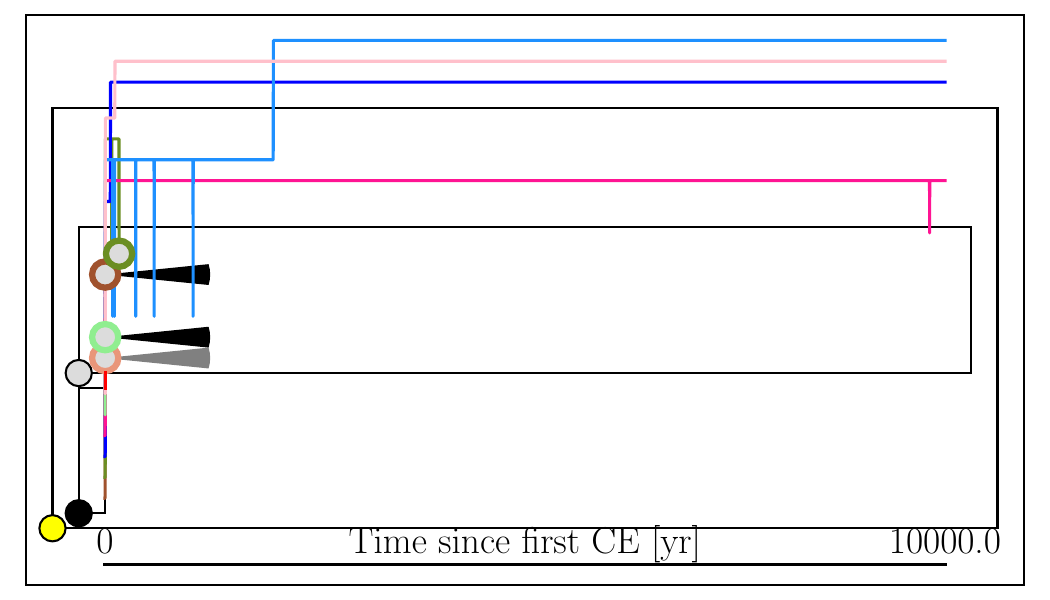}
    \caption{Further examples of the evolution of satellite systems, with the same symbols and lines as Figure~\ref{fig:example}. In addition, grey wedges mark the production of debris streams that are not bound to either planet (and are not counted in our statistics), while collisions without an attached wedge mark where a satellite tidally disrupts but the debris is on a collision course with the planet; both of these occur in the left-hand panel of the second row. The system shown in Fig~\ref{fig:example} and discussed in the main paper is also shown here in the middle of the fourth row.}
    \label{fig:appendix}
\end{figure*}

In Figure~\ref{fig:appendix}, we show further examples of the evolution of satellite systems, in the same manner as Figure~\ref{fig:example}. These are taken from the first N numbered systems in our simulation set, and hence form a random sample owing to the random initial conditions.

\bsp    % typesetting comment
\label{lastpage}

\end{document}